\documentclass[aps,prb,twocolumn,floatfix,superscriptaddress,showpacs,showkeys]{revtex4}
\usepackage{amsmath}
\usepackage{amsfonts}
\usepackage{graphicx}
\usepackage{amssymb}

\input{tcilatex}

\begin{document}

\title{The violation of the Hund's rule in semiconductor artificial atoms}
\author{C. F. Destefani}
\affiliation{Departamento de F\'{\i}sica, Universidade Federal de S\~{a}o Carlos,
13565-905 S\~{a}o Carlos-SP, Brazil}
\author{J. D. M. Vianna}
\affiliation{Instituto de F\'{\i}sica, Universidade de Bras\'{\i}lia, 70910-900 Bras\'{\i}%
lia-DF, Brazil}
\affiliation{Instituto de F\'{\i}sica, Universidade Federal da Bahia, 40210-340
Salvador-BA, Brazil}
\author{G. E. Marques}
\affiliation{Departamento de F\'{\i}sica, Universidade Federal de S\~{a}o Carlos,
13565-905 S\~{a}o Carlos-SP, Brazil}
\date{\today}

\begin{abstract}
The unrestricted Pople-Nesbet approach for real atoms is adapted to quantum
dots, the man-made artificial atoms, under applied magnetic field. Gaussian
basis sets are used instead of the exact single-particle orbitals in the
construction of the appropriated Slater determinants. Both system chemical
potential and charging energy are calculated, as also the expected values
for total and z-component in spin states. We have verified the validity of
the energy shell structure as well as the Hund's rule state population at
zero magnetic field. Above given fields, we have observed a violation of the
Hund's rule by the suppression of triplets and quartets states at the $1p$
energy shell, taken as an example. We also compare our present results with
those obtained using the $\mathbf{LS}$-coupling scheme for low electronic
occupations. We have focused our attention to ground-state properties for
GaAs quantum dots populated up to forty electrons.
\end{abstract}

\maketitle

The influence of spatial confinement on the physical properties such as the
electronic spectra of

0D structures is a topic of growing interest. Among such systems one could
select carriers and impurity atoms in metallic or semiconductor mesoscopic
structures,\cite{jask} as also atoms, ions and molecules trapped to
microscopic cavities,\cite{jask,conne,riv,waz,beekman}. In these systems,
the confinement becomes important whenever a quantum sizes equal the cavity
length. However, the energy spectrum of these systems is not only determined
by the spatial confinement and its geometrical shape, but also by
environmental facts such as electric and magnetic applied fields that break
or lower the general symmetries. Finally, many-body effects as
electron-electron interaction, may be even be more important than the
confinement itself. In any case, a correct description of physical
properties of the problem requires that the wavefunction must reflect both
the form of confinement and the appropriated boundary conditions.

Interesting confined systems are the semiconductor quantum dots structures
(QDs), also referred as man-made artificial atoms, built as low-dimensional
electronic gases when crystalline translation invariance is broken in all
three spatial dimensions and leading to discrete energy states, as in real
atoms. Various are the approaches that have been used to deal with
many-particle QDs. Among them, one can cite charging model,\cite{11,12,13,14}
correlated electron model,\cite{29} Green functions,\cite{30} Lanczos
algorithm,\cite{16} Monte Carlo method,\cite{31} Hartree-Fock (HF)
calculations,\cite{32,33,34,35} and density functional theory.\cite{36}

The charging model can reproduce well experimental findings for\ "metallic
dots". On the opposite side, with much lower electronic density, the
semiconductor dots requires a microscopic point of view to treat the
electron-electron interaction.

Here we will consider a QD defined by a "hard wall" spherical volume which
is appropriated for semiconductors grown inside glass matrices. Some of the
commonly studied topics are the formation of energy shells in their spectra,%
\cite{43} the control of their electronic correlations,\cite{44} the
formation of Wigner molecules,\cite{45} and the influence of the Coulomb
interaction in their spectra.\cite{46,47} In these spherically defined
artificial atoms both spin and orbital angular momenta are good quantum
numbers, and the low occupation many-particle eigenstates can be labelled
according to the $\mathbf{LS}$-coupling scheme,\cite{condon}. For occupation
number above four, the $\mathbf{LS}$-coupling scheme becomes extremely
cumbersome and, in this paper, we have chosen the unrestricted Pople-Nesbet
matrix approach\cite{szabo} of the single determinant self-consistent HF
formalism to treat shell configurations of dots containing up to forty
electrons, where we have calculated the total spin expected values, chemical
potential and charging energy. We show the changes induced by the magnetic
field on such approach, using a set of Gaussian basis (section II). Then we
discuss our main results (section III) by focusing on how magnetic field
determines the Zeeman splitting and induces violation of the Hund's rule.

Within the Unrestricted Hartree-Fock formalism (UHF), the $\alpha$ (spin-up)
and $\beta $ (spin-down) functions have different spatial components, $%
\chi_{i}(\mathbf{x})=\left\{\psi_{j}^{\alpha }(\mathbf{r})\alpha (\omega
),\psi _{j}^{\beta }(\mathbf{r})\beta (\omega )\right\}$, described by the
orbitals $\left\{\psi _{j}^{\alpha }|j=1...k\right\}$ ($\{\psi_{j}^{%
\beta}|j=1...k\}$). Therefore, an UHF wavefunction has the form $\left\vert
\Psi ^{UHF}\right\rangle =\left\vert \psi _{1}^{\alpha} \overline{\psi}%
_{1}^{\beta }...\right\rangle $, which represents open shells once no
spatial orbital can be doubly occupied. The closed shells can be also
obtained,\cite{szabo} more specifically, UHF functions are not necessarily
system eigenstates having well defined $L$ and $S$ values. Yet, the number
of carrier, $N$, must equal the the sum of spin-up and spin-down electrons,
as $N=N^{\alpha }+N^{\beta }$. The integration of the spin degrees of freedom%
\cite{szabo} yields two coupled HF equations that must be simultaneously
solved. They have the form $f^{\alpha /\beta }\left\vert \psi _{j}^{\alpha
/\beta }\right\rangle =\varepsilon _{j}^{\alpha /\beta }\left\vert \psi
_{j}^{\alpha /\beta }\right\rangle $, where the respective Fock operators
are given by 
\begin{equation}
f^{\alpha /\beta }=h_{j}+\sum_{a}^{N^{\alpha /\beta }}\left[J_{a}^{\alpha
/\beta }-K_{a}^{\alpha /\beta }\right] +\sum_{a}^{N^{\beta /\alpha
}}J_{a}^{\beta /\alpha }\text{.}  \label{fafb2}
\end{equation}%
Both $f^{\alpha }$ and $f^{\beta}$ include the kinetic($h_{j}$), the direct( 
$J_{a}^{\alpha /\beta}$) and the exchange ($K_{a}^{\alpha /\beta}$) terms
between electrons with same spin, and a direct term ($J_{a}^{\beta /\alpha}$
) between electrons with opposite spins. The interdependence among $%
f^{\alpha }$ ($f^{\beta }$) and $\psi _{j}^{\beta }$ ($\psi _{j}^{\alpha }$)
requires the simultaneous solution of the two HF equations. They yield the
sets $\left\{ \psi _{j}^{\alpha }\right\} $ and $\left\{ \psi _{j}^{\beta
}\right\} $ that should minimize the energy $E_{0}^{UHF}$ of the
unrestricted ground-state, $\left\vert \Psi _{0}^{UHF}\right\rangle $, given
by 
\begin{eqnarray}
E_{0}^{UHF} &=&\sum_{a}^{N^{\alpha }}h_{aa}^{\alpha }+\sum_{a}^{N^{\beta
}}h_{aa}^{\beta }+\frac{1}{2}\sum_{a}^{N^{\alpha }}\sum_{b}^{N^{\alpha }}%
\left[ J_{ab}^{\alpha \alpha }-K_{ab}^{\alpha \alpha }\right]  \notag \\
&&+\frac{1}{2}\sum_{a}^{N^{\beta }}\sum_{b}^{N^{\beta }}\left[ J_{ab}^{\beta
\beta }-K_{ab}^{\beta \beta }\right] +\sum_{a}^{N^{\alpha
}}\sum_{b}^{N^{\beta }}J_{ab}^{\alpha \beta }\text{.}  \label{eo1}
\end{eqnarray}%
In these expressions, $h_{aa}^{\alpha /\beta }=\left\langle \psi
_{a}^{\alpha /\beta }\right\vert h_{a}\left\vert \psi _{a}^{\alpha /\beta
}\right\rangle $, $J_{ab}^{\alpha \beta }=\left\langle \psi _{a}^{\alpha
}\right\vert J_{b}^{\beta }\left\vert \psi _{a}^{\alpha }\right\rangle
=\left\langle \psi _{b}^{\beta }\right\vert J_{a}^{\alpha }\left\vert \psi
_{b}^{\beta }\right\rangle =J_{ba}^{\beta \alpha }$, $J_{ab}^{\alpha \alpha
}=\left\langle \psi _{a}^{\alpha }\right\vert J_{b}^{\alpha }\left\vert \psi
_{a}^{\alpha }\right\rangle =\left\langle \psi _{b}^{\alpha }\right\vert
J_{a}^{\alpha }\left\vert \psi _{b}^{\alpha }\right\rangle =J_{ba}^{\alpha
\alpha }$ (with an analog term for $J_{ab}^{\beta \beta }$), and $%
K_{ab}^{\alpha \alpha }=\left\langle \psi _{a}^{\alpha }\right\vert
K_{b}^{\alpha }\left\vert \psi _{a}^{\alpha }\right\rangle =\left\langle
\psi _{b}^{\alpha }\right\vert K_{a}^{\alpha }\left\vert \psi _{b}^{\alpha
}\right\rangle =K_{ba}^{\alpha \alpha }$ (with an analog term for $%
K_{ab}^{\beta \beta }$).

The Pople-Nesbet approach transforms the UHF equations into a matrix
formulation by expanding $\psi _{i}^{\alpha /\beta }$ in a set of known
basis functions $\left\{ \phi _{\nu }|\nu =1...k\right\} $, 
\begin{equation}
\psi _{i}^{\alpha /\beta }=\sum_{\nu }C_{\nu i}^{\alpha /\beta }\phi _{\nu }%
\text{,}  \label{exsp}
\end{equation}%
where the expansion coefficients $C_{\nu i}^{\alpha /\beta }$ become the
parameters to be iterated. When Eq.~\ref{exsp} is inserted into Eq. \ref%
{fafb2} one obtains the $k$ $\times $ $k$ coupled matrix equations, 
\begin{equation}
\mathbf{F}^{\alpha /\beta }\mathbf{C}^{\alpha /\beta }=\mathbf{SC}^{\alpha
/\beta }\mathbf{\varepsilon }^{\alpha /\beta }\text{,}  \label{pn}
\end{equation}%
where $\mathbf{S}$ is the positive defined overlap ($S_{\mu \nu
}=\left\langle \phi _{\mu }\right. \left\vert \phi _{\nu }\right\rangle $)
between basis functions, $\mathbf{C}^{\alpha /\beta }$ are the expansion
coefficient matrices whose columns describe each spatial orbital $\psi
_{i}^{\alpha /\beta }$, $\mathbf{\varepsilon }^{\alpha /\beta }$ are the
diagonal matrices of the orbital energies ($\varepsilon _{i}^{\alpha /\beta
} $), and $\mathbf{F}^{\alpha /\beta }$ are the Fock operators, 
\begin{equation}
F_{\mu \nu }^{\alpha /\beta }=\int d^{3}\mathbf{r}\phi _{\mu }^{\ast }(%
\mathbf{r})f^{\alpha /\beta }\phi _{\nu }(\mathbf{r})\text{.}  \label{irres}
\end{equation}%
At this point becomes convenient to introduce the charge density for the
spin-up and -down electrons, defined as $\rho ^{\alpha /\beta }(\mathbf{r}%
)=\sum_{a}^{N^{\alpha /\beta }}\left\vert \psi _{a}^{\alpha /\beta }(\mathbf{%
r})\right\vert ^{2}=\sum_{\mu }\sum_{\nu }P_{\mu \nu }^{\alpha /\beta }\phi
_{\mu }(\mathbf{r})\phi _{\nu }^{\ast }(\mathbf{r})$, where the elements of
the respective density matrices are $P_{\mu \nu }^{\alpha /\beta
}=\sum_{a}^{N^{\alpha /\beta }}C_{\mu a}^{\alpha /\beta }C_{\nu a}^{\alpha
/\beta \ast }$. Thus, one can define two new quantities: i) The total charge
density, $\rho ^{T}(\mathbf{r})=\rho ^{\alpha }(\mathbf{r})+\rho ^{\beta }(%
\mathbf{r})$, that determines $N$ when integrated over all space; ii) The
spin density, $\rho ^{S}(\mathbf{r})=\rho ^{\alpha }(\mathbf{r})-\rho
^{\beta }(\mathbf{r})$, that yields $2M_{S}$ after integration over all
space. The unrestricted wavefunctions are eigenfunctions of $S_{Z}$, but not
necessarily of $S^{2}$, therefore one can define the total charge ( $\mathbf{%
P}^{T}=\mathbf{P}^{\alpha }+\mathbf{P}^{\beta }$) and the spin ( $\mathbf{P}%
^{S}=\mathbf{P}^{\alpha }-\mathbf{P}^{\beta }$) density matrices for the
system. The elements of the two Fock matrices are obtained as $F_{\mu \nu
}^{\alpha /\beta }=T_{\mu \nu }+G_{\mu \nu }^{\alpha /\beta }$, where $%
T_{\mu \nu }=-\hbar ^{2}/(2m)\left\langle \phi _{\mu }\right\vert \nabla
^{2}\left\vert \phi _{\nu }\right\rangle $ and $G_{\mu \nu }^{\alpha /\beta
}=e^{2}/\varepsilon \sum_{\lambda }\sum_{\sigma }[P_{\lambda \sigma
}^{T}\left\langle \phi _{\mu }\phi _{\sigma }\right\vert \left\vert \mathbf{r%
}_{1}-\mathbf{r}_{2}\right\vert ^{-1}\left\vert \phi _{\nu }\phi _{\lambda
}\right\rangle -P_{\lambda \sigma }^{\alpha /\beta }\left\langle \phi _{\mu
}\phi _{\sigma }\right\vert \left\vert \mathbf{r}_{1}-\mathbf{r}%
_{2}\right\vert ^{-1}\left\vert \phi _{\lambda }\phi _{\nu }\right\rangle ]$%
, with $\varepsilon $ being the material dielectric constant. \qquad\ 

The self-consistency lies in the fact that both $\mathbf{F}$ and $\mathbf{P}$
depend on $\mathbf{C}$, while the coupling of spin-up and -down equations
occurs since $\mathbf{F}^{\alpha }$ ($\mathbf{F}^{\beta }$) depends on $%
\mathbf{P}^{\beta }$ ($\mathbf{P}^{\alpha }$) through $\mathbf{P}^{T}$.

The procedure to solve Eq. \ref{pn} is: i) Given a confinement potential,
one specifies $N$ and $\left\{ \phi _{\mu }\right\} $; ii) The integrations
on $S_{\mu \nu }$ and $T_{\mu \nu }$ are performed; iii) An initial guess is
used for $\mathbf{P}^{\alpha /\beta }$ and $\mathbf{P}^{T}$, perform
two-electron integrals for $\mathbf{G}^{\alpha /\beta }$ and $\mathbf{T}$ to
construct $\mathbf{F}^{\alpha /\beta }$. $\ $\ Diagonalize $\mathbf{F}%
^{\alpha /\beta }$ to get $\mathbf{C}^{\alpha /\beta }$ and $\mathbf{%
\varepsilon }^{\alpha /\beta }$, and form new $\mathbf{P}^{\alpha /\beta }$;
iv) This iteration is repeated until the desired convergence for $%
E_{0}^{UHF} $ is reached. The Pople-Nesbet ground-state energy is 
\begin{equation}
E_{0}^{UHF}=\frac{1}{2}\sum_{\mu }\sum_{\nu }\left[ P_{\nu \mu }^{T}T_{\mu
\nu }+P_{\nu \mu }^{\alpha }F_{\mu \nu }^{\alpha }+P_{\nu \mu }^{\beta
}F_{\mu \nu }^{\beta }\right] \text{ .}  \label{ea_U}
\end{equation}

The unrestricted functions are not, in general, eigenstates of $S^{2}$, thus
we calculate the spin expectation values as \cite{szabo} 
\begin{eqnarray}
\left\langle S^{2}\right\rangle _{UHF} &=&\left( \frac{N^{\alpha }-N^{\beta }%
}{2}\right) \left( \frac{N^{\alpha }-N^{\beta }}{2}+1\right) +N^{\beta } 
\notag \\
&&-\sum_{a}^{N^{\alpha }}\sum_{b}^{N^{\beta }}\left[ \sum_{\mu }\sum_{\nu
}C_{\mu a}^{\alpha \text{ }\ast }C_{\nu b}^{\beta }S_{\mu \nu }\right] ^{2}%
\text{,}  \label{S2}
\end{eqnarray}%
and 
\begin{equation}
\left\langle S_{Z}\right\rangle _{UHF}=\frac{1}{2}\sum_{\mu }\sum_{\nu
}\left( P_{\nu \mu }^{\alpha }-P_{\nu \mu }^{\beta }\right) S_{\mu \nu }%
\text{.}  \label{SZ}
\end{equation}

As an application of the Pople-Nesbet approach, we consider a QD with radius 
$R_{0}$ confined to an infinite spherical potential in the presence of a
magnetic field $\mathbf{B}=(0,0,B_{0})$ and populated up to forty electrons.
Its single-particle Hamiltonian is 
\begin{equation}
H_{0}=\frac{\hbar ^{2}}{2m}\left( \mathbf{k}+\frac{e}{\hbar c}\mathbf{A}%
\right) ^{2}+g\frac{\mu _{B}}{\hbar }\mathbf{B}\cdot \mathbf{S}\text{,}
\label{H1P}
\end{equation}%
where $\mu _{B}$ is the Bohr magneton, $g$ is the bulk $g$-factor, and we
use the symmetric gauge, $\mathbf{A}=(\mathbf{B}\times \mathbf{r})/2$. Using
atomic units, $E_{Ry}=e^{2}/(2a_{0})$(Rydberg) for the energy, and $%
a_{0}=\hbar ^{2}/(m_{0}e^{2})$(Bohr radius) for length, the Hamiltonian $%
H_{0}$ can be written in dimensionless form 
\begin{eqnarray}
H_{0} &=&\frac{1}{\widetilde{m}}\frac{a_{0}^{2}}{R_{0}^{2}}\left[ -\frac{1}{%
x^{2}}\frac{\partial }{\partial x}\left( x^{2}\frac{\partial }{\partial x}%
\right) +\frac{\mathbf{L}^{2}}{x^{2}}\right.  \label{Ha0} \\
&&\left. +\frac{R_{0}^{2}}{2l_{B}^{2}}\left( L_{Z}+\widetilde{m}%
gS_{Z}\right) +\frac{R_{0}^{4}}{4l_{B}^{4}}x^{2}\sin ^{2}(\theta )\right] 
\text{,}  \notag
\end{eqnarray}%
where $\widetilde{m}=m/m_{0}$, $l_{B}=\sqrt{\hbar c/(eB_{0})}$ is the
magnetic length, and $x=r/R_{0}$. Without magnetic field, the normalized
spatial eigenfunctions of $H_{0}$ are given by 
\begin{equation}
\phi _{\nu }(\mathbf{r})=\left[ \frac{2}{R_{0}^{3}}\frac{1}{\left[
j_{l+1}\left( \alpha _{nl}\right) \right] ^{2}}\right] ^{1/2}j_{l}\left(
\alpha _{nl}x\right) Y_{l,m_{l}}(\theta ,\phi )\text{.}  \label{est1p}
\end{equation}

The boundary condition at the surface, $r=R_{0}$ (or $x=1$), determines $%
\alpha _{nl}$ as the $n^{th}$ zero of the spherical Bessel function $%
j_{l}(\alpha _{nl}x)$ and $Y_{l,m_{l}}(\theta ,\phi )$ is the spherical
harmonic.

The Hamiltonian for the electron-electron interaction in atomic units
becomes $H_{ee}=(a_{0}/R_{0})2/(\varepsilon \left\vert \mathbf{x}_{1}-%
\mathbf{x}_{2}\right\vert )$, where the usual multipole expansion for $|%
\mathbf{x}_{1}-\mathbf{x}_{2}|^{-1}$ is used in our calculations.

The spatial orbitals in $\psi _{i}^{\alpha /\beta }$ define the six lowest
energy shells ($1s$, $1p$, $1d$, $2s$, $1f$, $2p$) without magnetic field,%
\cite{walecka}. Thus, the index $\nu \equiv n,l,m_{l}$ can assume up to $40$
($20$ spin-up and $20$ -down) possible values for those shells. Certainly,
the magnetic field lifts both spin and orbital degeneracies. Let us consider
a GaAs dot, a wide-gap semiconductor having $\widetilde{m}=0.065$, $g=0.45$
and $\varepsilon =12.65$.

The inclusion of a magnetic field requires modifications on the UHF
equations. The spin-independent linear and quadratic magnetic terms are
easyly added to the definitions of both $h$ in $f^{\alpha /\beta }$ and $%
T_{\mu \nu }$. However, the inclusion of the spin-dependent linear term ($%
\sim B_{0}S_{Z}$) to $h$ in $f^{\alpha /\beta }$ requires decomposition of
kinetic and Coulomb terms into $T_{\mu \nu }^{\alpha /\beta }$ and $G_{\mu
\nu }^{\alpha /\beta }$. Thus, under a magnetic field we should make the
substitution $P_{\nu \mu }^{T}T_{\mu \nu }\Rightarrow P_{\nu \mu }^{\alpha
}T_{\mu \nu }^{\alpha }+P_{\nu \mu }^{\beta }T_{\mu \nu }^{\beta }$ in $%
E_{0}^{UHF}$ (Eq.~\ref{ea_U}).

\begin{figure}[tbp]
\includegraphics[width=1.0\linewidth]{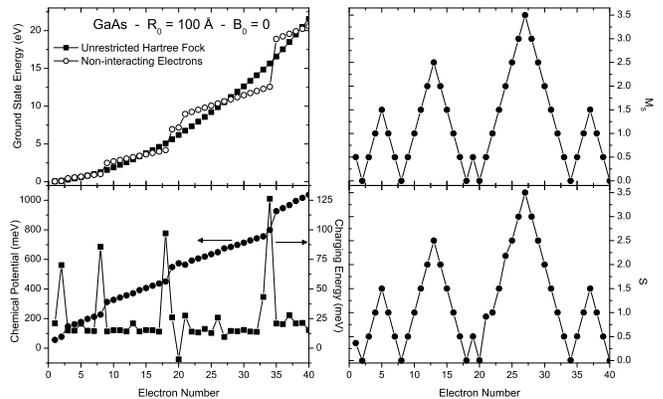} 
\caption{Unrestricted ground-state energy for a $R_{0}=100$ \AA\ GaAs QD
without magnetic field. In the upper left panel we compare the unrestricted
and the non-interacting energies, where the QD energy shell structure is
visible. The bottom left panel shows QD chemical potential (left scale) and
charging energy (right scale). The former displays abrupt change always when
a new shell starts to be populated, while the latter presents larger
(smaller) peaks when a shell is fully filled (half-filled), a direct
consequence of the Hund's rule. The bottom and upper right panels show,
respectively, the $N$-evolution of the expected values of total spin and its
z-projection.}
\label{fig1}
\end{figure}

Another important detail refers to the orbital basis $\left\{ \phi _{\nu
}|\nu =1...k\right\} $. Instead of the exact spherical Bessel functions of
Eq.\ref{est1p}, the radial part of each orbital was decomposed in a sum
involving five Gaussians confined to region $x\leq 1$, while the angular
part is maintained as defined by its symmetry, as 
\begin{eqnarray}
\phi _{n,l,m_{l}}(x,\theta ,\phi ) &=&N_{nl}\left( 1-x\right)
^{n}x^{l}\prod_{i=1}^{n-1}\left( \widetilde{\alpha }_{il}-x\right)
\label{gaus} \\
&&\times \sum_{k=1}^{5}V_{k}e^{-D_{k}R_{0}^{2}x^{2}}Y_{l,m_{l}}(\theta ,\phi
)\text{,}  \notag
\end{eqnarray}%
where $N_{nl}$ is the normalization, the polynomial $\left( 1-x\right) ^{n}$
garanties the boundary $x=1$, the polynomial in $x^{l}$ is required for $l>0$
states at the origin $x=0$, the product in $\left( \widetilde{\alpha }%
_{il}-x\right) $ nulls functions at the zeros $\widetilde{\alpha }_{il}$ of
the respective spherical Bessel function transposed to the interval $0\leq
x\leq 1$, and the last sum involves an expansion into $5$ Gaussians. Higher
order expansion did not show any improvement for $N\leq 40$. The Gaussian
coefficients $V_{k}$ and exponents $D_{k}$ are determined for each value of $%
R_{0}$, by maximizing the superposition between Eqs. \ref{est1p} and \ref%
{gaus}. Once $V_{k}$ and $D_{k}$ are determined, we run our UHF code for
each value of $R_{0}$ and $N$ , and find the parameters $C_{\nu i}^{\alpha
/\beta }$ that better describe Eq. \ref{exsp} and give the minimal energy in
Eq. \ref{ea_U}.

At last, we have calculated two quantities that will be used in the
description of our results. The first one is the QD chemical potential,
which yields the energy difference between two successive ground states, 
\begin{equation}
\mu _{dot}(N)=E_{0}(N)-E_{0}(N-1)\text{ .}  \label{pqui}
\end{equation}%
The second one is the QD charging energy, which yields the energy cost to
add an extra electron to the system, 
\begin{equation}
E_{char}(N)=E_{0}(N+1)-2E_{0}(N)+E_{0}(N-1)\text{ .}  \label{ecar}
\end{equation}%
From these two last equations, one can also see that $E_{char}(N)=\mu
_{dot}(N+1)-\mu _{dot}(N)$.

We show in Fig. \ref{fig1} the results of a UHF Pople-Nesbet calculation for
a GaAs QD having $R_{0}=100$ \AA , at zero magnetic field. In the left upper
panel we have compared the non-interacting electron problem and the UHF
results as function of occupation. The shell structure occurs for magic
numbers $N=2$, $8$, $18$, $20$, $34$ and $40$. It is observed that
electron-electron interaction decreases (increases) the non-interacting
ground-state energy when occupation corresponds to a shell less (more) than
half-filled. At exactly half-filled cases $N=5$, $13$, $27$ and $37$, the
interacting and non-interacting energy values are approximately equal.

In the bottom left panel of Fig.~\ref{fig1} we show both QD chemical
potential (left scale, Eq.~\ref{pqui}) and charging energy (right scale, Eq.~%
\ref{ecar}), where the respective values of $E_{0}$ were obtained from the
unrestricted calculation presented in the left upper panel. Notice that $\mu
_{dot}$ linearly increases as the occupation increases inside a given shell.
When such shell is totally filled, there is an abrupt change in $\mu _{dot}$
indicating that the following shell starts its occupation. Also, the higher
the occupation, the more abrupt is this change. An anomalous behavior seems
to occur with the $2s$ shell, whose $\mu _{dot}$-value is larger than the $1f
$ shell, that has higher energy. The charging energy is another form to
verify not only the presence of shell strusture in the spectrum, but also
the validity of Hund's rule for the filling of such shells. In principle, $%
E_{char}$ must present larger (smaller) peaks when the total (half)
occupation of a shell is achieved. The first fact is due to the higher
difficulty in adding an electron to a QD in a filled shell state, while the
second one refers to Hund's rule, which states that electrons must be added
to the system with their spin being parallel, until all possible orbitals
inside a given shell be occupied, making the total energy of the system be
decreased because of the maximized exchange contribution. However, some
discrepancies are verified in $E_{char}$: the smaller peak of $N=27$ occurs
here at $N=26$, and the larger peak of $N=20$ is negative. 
\begin{figure}[tbp]
\includegraphics[width=1.0\linewidth]{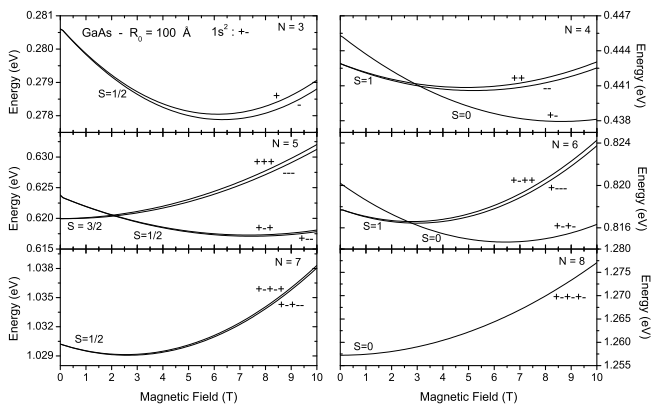} 
\caption{Violation of Hund's rule induced by magnetic field in the $R_{0}=100
$ \AA\ GaAs QD of the previous figure. The panels show the successive
occupation (indicated in the upper right corner of each panel) of the $1p$
shell, assuming that the $1s$ shell remains populated by one spin-up and one
spin-down electron. The possible spin configurations for given $N$ are
indicated by $+$ (spin-up) and $-$ (spin-down). For $B_{0}=0$ the spin
sequence is $1/2-1-3/2-1-1/2-0$, while at fields higher than $3$ T, it
changes to $1/2-0-1/2-0-1/2-0$. It interesting to observe that a magnetic
field can suppress the energy triplets and quartets from the QD spectrum.}
\label{fig2}
\end{figure}

The right bottom and upper panels of Fig.\ref{fig1} show respectively the $N$%
-evolution of the total spin $S$ and its projection $M_{S}$ as calculated
from Eqs.~\ref{S2} and \ref{SZ} for the unrestricted energies. Notice that,
with no magnetic field, Hund's rule seems to be followed; the $M_{S}$
expected value oscillates from $0$ in a filled shell to its maximum in a
half-filled shell, when it starts to decrease again on the way to the
closing of the shell; the maxima are $M_{S}=1/2$, $3/2$, $5/2$ and $7/2$ for 
$s$, $p$, $d$ and $f$ shells, respectively. The $S$ expected value yielded
by the unrestricted formalism is also very reasonable; discrepancies are
only observed at $N=24$, where $S>2$, and at $N=21$, where $S>1/2$. We
believe that both discrepancies related to the $2s$ shell or to its
surroundings - $\mu _{dot}$ larger than the one of $1f$ shell, negative peak
for $N=20$ in $E_{char}$, and almost doubled $S$ expected value for $N=21$ -
are caused by the non-reasonable Gaussian reproduction of this orbital.\cite%
{artigo}

By focusing on the $1p$ shell we show in Fig. \ref{fig2}, for the same QD of
the previous figure, how a finite magnetic field is able to violate Hund's
rule in the system. Panels from left to right and from up to bottom show the
successive ground state energies from $N=3$ to $N=8$ as this shell is
filled, always considering that the $1s$ shell remain fully occupied by two
electrons, one up and one down; the distinct possible spin configurations
for each $N$ are indicated by $+$ (up) and $-$ (down). In addition to the
small Zeeman effect present in all occupations, there is a changing of
ground state spins at $N=4$, $5$ and $6$ as the field is increased. Notice
that at zero field the spin sequence is $1/2-1-3/2-1-1/2-0$; in a field
above $3$ T it becomes $1/2-0-1/2-0-1/2-0$, meaning that quartets and
triplets are suppressed by the magnetic field, and the ground state of the
system starts to oscillate only between singlets and doublets at high fields
as $N$ increases. When this $1p$ shell is half-filled ($N=5$), the ground
state goes from a quartet to a doublet at $B_{0}\simeq 2$ T; when it has one
electron more ($N=6$) or less ($N=4$) than that, it goes from a triplet to a
singlet at $B_{0}\simeq 3$ T.

At last, in order to prove the efficiency of the Pople-Nesbet approach, we
compared the results from this UHF self-consistent matrix formulation with
the ones obtained from the $\mathbf{LS}$-coupling scheme used in Ref.~[%
\onlinecite{LS}], where a GaAs QD having $R_{0}=90$ \AA\ was considered, and
the quadratic term in $B_{0}$ was neglected since only small fields were
considered; also, only $N=2$ and $N=3$ occupations were calculated, since
the states were exactly built (not only a single Slater determinant), and
the electron-electron interaction was included by using perturbation theory,
justified at such radius. At zero field the energies for $N=2$ are $16.5$
meV ($\mathbf{LS}$) and $16.1$ meV (UHF), while for $N=3$ they are $34.8$
meV ($\mathbf{LS}$) and $33.9$ meV\ (UHF); so, the formalism here used
indeed give smaller ground state energies than the $\mathbf{LS}$
perturbation scheme. We have also checked the validity of neglecting the
quadratic term in $B_{0}$ for fields smaller than $2$ T. One should
emphasize that a disadvantage of the UHF approach is that, in principle, it
is not sure that one gets trustable information about the $L$ and $S$
expected values of QD states; on the other hand, the applicability of the $%
\mathbf{LS}$ scheme is highly decreased as the QD occupation increases.

We have shown how the unrestricted Pople-Nesbet approach applied to a
spherical QD under a magnetic field yields a reasonable description of its
energetic spectrum, where a maximum occupation of $40$ electrons has been
considered. We have seen how both QD chemical potential and charging energy
reproduce the filling and half-filling structures of its energy shells at
zero field. With the total spin expected value for each occupation in a
given radius, we have seen that the Hund rule is satisfied at zero field;
however, under a finite field, we have shown that it is violated and, at
given values of the field which depend on QD parameters, transitions that
change a given ground state symmetry are observed.

We acknowledge support from FAPESP-Brazil. 

\end{document}